\documentclass[11pt]{article}
\usepackage{amssymb,amsfonts}
\usepackage{graphics,psboxit,amsmath}
\usepackage{subfigure}
\usepackage{graphicx}
\usepackage{verbatim}
\usepackage{color}
\usepackage{hyperref}
\hypersetup{colorlinks}

\definecolor{darkred}{rgb}{1,0,0}
\definecolor{darkgreen}{rgb}{0,0.5,0}
\definecolor{darkblue}{rgb}{0,0,1}
\definecolor{orange}{rgb}{1,0.5,0}
\definecolor{green}{rgb}{0,1,0}
\definecolor{purple}{rgb}{.5,0,1}

\hypersetup{ colorlinks,
linkcolor=darkred,
filecolor=darkgreen,
urlcolor=darkblue,
citecolor=darkblue,
linktocpage=true }

\definecolor{markcolor}{rgb}{.25,0,1}

\definecolor{markcolor2}{rgb}{1,0,0}

\definecolor{markcolor3}{rgb}{0,1,0}

\def\hybrid{\topmargin -10pt    \oddsidemargin 0.1in %%%%%%%%%%%%%% Archive-30pt
        \headheight 0pt \headsep 0pt
        \textwidth 15.0cm      % A4 paper
        \textheight 21.15cm       % A4 paper
        \marginparwidth .875in
        \parskip 5pt plus 1pt   \jot = 1.5ex}

\hybrid

\catcode`\@=11

\def\marginnote#1{}
\newcount\hour
\newcount\minute
\newtoks\amorpm
\hour=\time\divide\hour by60 \minute=\time{\multiply\hour by60
\global\advance\minute by-\hour}
\edef\standardtime{{\ifnum\hour<12 \global\amorpm={am}%
        \else\global\amorpm={pm}\advance\hour by-12 \fi
        \ifnum\hour=0 \hour=12 \fi
        \number\hour:\ifnum\minute<10 0\fi\number\minute\the\amorpm}}
\edef\militarytime{\number\hour:\ifnum\minute<10 0\fi\number\minute}
\def\draftlabel#1{{\@bsphack\if@filesw {\let\thepage\relax
   \xdef\@gtempa{\write\@auxout{\string
      \newlabel{#1}{{\@currentlabel}{\thepage}}}}}\@gtempa
   \if@nobreak \ifvmode\nobreak\fi\fi\fi\@esphack}
        \gdef\@eqnlabel{#1}}
\def\@eqnlabel{}
\def\@vacuum{}
\def\draftmarginnote#1{\marginpar{\raggedright\scriptsize\tt#1}}

\def\draft{\oddsidemargin -.5truein
        \def\@oddfoot{\sl preliminary draft \hfil
        \rm\thepage\hfil\sl\today\quad\militarytime}
        \let\@evenfoot\@oddfoot \overfullrule 3pt
        \let\label=\draftlabel
        \let\marginnote=\draftmarginnote
   \def\@eqnnum{(\theequation)\rlap{\kern\marginparsep\tt\@eqnlabel}%
\global\let\@eqnlabel\@vacuum}  }

\def\draft2{
        \def\@oddfoot{\sl preliminary draft \hfil
        \rm\thepage\hfil\sl\today\quad\militarytime}
        \let\@evenfoot\@oddfoot \overfullrule 3pt
        \let\label=\draftlabel
        \let\marginnote=\draftmarginnote
   \def\@eqnnum{(\theequation)\rlap{\kern\marginparsep\tt\@eqnlabel}%
\global\let\@eqnlabel\@vacuum}  }

\def\preprint{\twocolumn\sloppy\flushbottom\parindent 2em
        \leftmargini 2em\leftmarginv .5em\leftmarginvi .5em
        \oddsidemargin -.5in    \evensidemargin -.5in
        \columnsep .4in \footheight 0pt
        \textwidth 10.in        \topmargin  -.4in
        \headheight 12pt \topskip .4in
        \textheight 6.9in \footskip 0pt
        \def\@oddhead{\thepage\hfil\addtocounter{page}{1}\thepage}
        \let\@evenhead\@oddhead \def\@oddfoot{} \def\@evenfoot{} }

\def\numberbysection{\@addtoreset{equation}{section}
        \def\theequation{\thesection.\arabic{equation}}}

\def\underline#1{\relax\ifmmode\@@underline#1\else
        $\@@underline{\hbox{#1}}$\relax\fi}

\def\titlepage{\@restonecolfalse\if@twocolumn\@restonecoltrue\onecolumn
     \else \newpage \fi \thispagestyle{empty}\c@page\z@
        \def\thefootnote{\fnsymbol{footnote}} }

\def\endtitlepage{\if@restonecol\twocolumn \else \newpage \fi
        \def\thefootnote{\arabic{footnote}}
        \setcounter{footnote}{0}}  %\c@footnote\z@ }

\catcode`@=12 \relax

\def\figcap{\section*{Figure Captions\markboth
        {FIGURECAPTIONS}{FIGURECAPTIONS}}\list
        {Figure \arabic{enumi}:\hfill}{\settowidth\labelwidth{Figure
999:}
        \leftmargin\labelwidth
        \advance\leftmargin\labelsep\usecounter{enumi}}}
 \relax
\def\tablecap{\section*{Table Captions\markboth
        {TABLECAPTIONS}{TABLECAPTIONS}}\list
        {Table \arabic{enumi}:\hfill}{\settowidth\labelwidth{Table
999:}
        \leftmargin\labelwidth
        \advance\leftmargin\labelsep\usecounter{enumi}}}
 \relax
\def\reflist{\section*{References\markboth
        {REFLIST}{REFLIST}}\list
        {[\arabic{enumi}]\hfill}{\settowidth\labelwidth{[999]}
        \leftmargin\labelwidth
        \advance\leftmargin\labelsep\usecounter{enumi}}}
 \relax
\makeatletter
\newcounter{pubctr}
\def\publist{\@ifnextchar[{\@publist}{\@@publist}}
\def\@publist[#1]{\list
        {[\arabic{pubctr}]\hfill}{\settowidth\labelwidth{[999]}
        \leftmargin\labelwidth
        \advance\leftmargin\labelsep
        \@nmbrlisttrue\def\@listctr{pubctr}
        \setcounter{pubctr}{#1}\addtocounter{pubctr}{-1}}}
\def\@@publist{\list
        {[\arabic{pubctr}]\hfill}{\settowidth\labelwidth{[999]}
        \leftmargin\labelwidth
        \advance\leftmargin\labelsep
        \@nmbrlisttrue\def\@listctr{pubctr}}}
 \relax
\makeatother

\def\be{\begin{equation}}
\def\ee{\end{equation}}
\def\ba{\begin{eqnarray}}
\def\ea{\end{eqnarray}}

\def\no{\noindent}

\def\IR{\relax{\rm I\kern-.18em R}}

\def\bse{\begin{small}\begin{equation*}}
\def\ese{\end{equation*}\end{small}}

\begin{document}
%\draft2

%\renewcommand{\theequation}{\arabic{equation}}
%\renewcommand{\theequation}{\thesection.\arabic{equation}}

\renewcommand{\theequation}{\thesection.\arabic{equation}}
\csname @addtoreset\endcsname{equation}{section}

\newcommand{\eqn}[1]{(\ref{#1})}

\begin{titlepage}
\begin{center}
\strut\hfill
\vskip 1.3cm

%\hfill  [hep-th]\\

\vskip .5in

{\large \bf Classical integrable defects as quasi B\"{a}cklund transformations}

\vskip 0.5in

{{\bf Anastasia Doikou}} \vskip 0.2in

 \vskip 0.02in
{\footnotesize
 Department of Mathematics, Heriot-Watt University,\\
EH14 4AS, Edinburgh, United Kingdom}
\\[2mm]

\vskip .1cm

%\vskip -.15in

{\footnotesize {\tt E-mail: a.doikou@hw.ac.uk}}\\

\end{center}

\vskip 1.0in

\centerline{\bf Abstract}
We consider the algebraic setting of classical defects in discrete and continuous integrable theories. We derive the ``equations of motion'' on the defect point via the space-like and time-like description. We then exploit the structural similarity of these equations with the discrete and continuous B\"{a}cklund transformations. And although these equations are similar they are not exactly the same to the B\"{a}cklund transformations. We also consider specific examples of integrable models to demonstrate our construction, i.e. the Toda chain and the sine-Gordon model. The equations of the time (space) evolution of the defect (discontinuity) degrees of freedom for these models are explicitly derived.
\no

\vfill

\end{titlepage}
\vfill \eject

%\def\baselinestretch{1.2}
%\baselineskip 10 pt
%\noindent

\tableofcontents

\section{Introduction}

The issue of  quantum and classical integrable defects has been the subject of consistently increasing research interest in recent years \cite{delmusi}--\cite{robertson}. Many results from various perspectives were produced for both quantum and classical integrable models and many interesting interconnections were pointed out. For instance the elucidation of the local defect as a ``frozen'' B\"{a}cklund transformation \cite{corrigan-atft, corrigan-NLS} was a significant observation providing a novel way of understanding local defects for integrable classical field theories. And although the interpretation of the defect matrix as a``frozen'' B\"{a}cklund transformation is a notable piece of information, a systematic Hamiltonian/algebraic description of this notion has been hitherto missing, and this is precisely the main objective in this paper.

More specifically, we shall introduce here the notion of {\it quasi B\"{a}cklund} transformations for both discrete and continuous classical integrable models in the presence of local defects. This concept is essentially associated to the local equations of motion around the defect point. Note that in the usual B\"{a}cklund transformation formulation one deals with a space and a time differential (difference) equation, and these two equations are simultaneously satisfied. In the presence of local defects however this is not the case, as the space and time differential (difference) equations that are obtained are not satisfied simultaneously. It is worth noting that these two equations in the frame of defects arise  as local equations of motion around the defect point (more details on this will be presented later in the text, see also \cite{avan-doikou}).
In the discrete case in particular one explicitly derives the time components of the Lax pair around the defects point, and these turn out to be slightly ``deformed'' compared to the bulk quantities. In the continuous case on the other hand due to imposed analyticity conditions around the defect point the time components of the Lax pair coincide with the left/right bulk theory quantities (for more details we refer the interested reader to \cite{avan-doikou}). These connections are of course remarkable and a detailed analysis is given in this article, however a deeper understanding of the particular solutions of these equations for various physically relevant models should be further pursued (see some relevant discussion in \cite{aguire}).

Let us briefly outline the content on this paper: in the next section we briefly review the auxiliary linear problem for both discrete and continuous classical integrable models. We also recall the Darboux-B\"{a}cklund transformations as suitable gauge transformations that leave the auxiliary linear problem invariant. The underlying classical algebras associated to the Lax pair are briefly described, and the notion of the ``dual'' formulation introduced in \cite{ACDK} is also presented. A note on the algebraic content of the B\"{a}cklund transformations is also given based mainly on the fact that the auxiliary functions are versions of classical vertex operators \cite{doikou15}.
In section 3 we focus on the problem under consideration i.e. the description of the local classical defects as quasi B\"{a}cklund transformations. More precisely, we discuss the equations of motion associated to the defect degrees of freedom, and ascertain their resemblance to the $t$ part of the B\"{a}cklund transformation. We also consider defects along the $t$ axis (``dual'' description) and thus we are able to derive the $x$ part of the B\"{a}cklund transformation. Jump conditions on the fields and their time derivatives are also obtained similar to the ones along the $x$ direction, and are part of the $x$ B\"{a}cklund transformation conditions. The underlying algebra for the Lax matrices as well as the time and space defect matrices are also discussed.

\section{The general setting}

We shall review in this section the main setting in describing continuous and discrete integrable classical models. The formulation we pursue here is based on two main building blocks, i.e. the auxiliary linear algebra on the one hand and the existence of an underlying classical algebra. We shall describe below both semi and fully discrete integrable models as well as continuous integrable theories. Moreover, we shall review the Darboux-B\"{a}cklund transformations as suitable gauge transformation associated to the auxiliary linear problem.

\subsection{Auxiliary linear problem $\&$ Darboux-B\"{a}cklund transformations}

\subsubsection{The discrete case}

We begin our brief review on B\"{a}cklund transformations in the context of the auxiliary linear problem considering semi-discrete integrable models, i.e. models with discrete space and continuous time parameters. We shall then briefly review fully discrete models, and in the next subsection we shall discuss the continuous space time case (see also e.g. \cite{wahlq, matveev, Backlund-book} and references therein).

The auxiliary linear problem in the semi-discrete case is expressed via the Lax pair $\Big(L_n(t; \lambda),\ V_n(t;\lambda)\Big)$ satisfying the following set of equations:
\ba
&&\Psi_{n+1}(t;\lambda) = L_n(t;\lambda)\ \Psi_n(t;\lambda)\cr
&& {\partial \Psi_n(t;\lambda) \over \partial t} = V_n(t;\lambda)\ \Psi_n(t;\lambda) \label{set1}
\ea
$L,\ V$ are in general ${\cal N}\times {\cal N}$ matrices depending on fields (algebraic objects) and maybe on the spectral parameter $\lambda$; $\Psi$ is the auxiliary vector function, which may be also thought of as a classical vertex operator (see \cite{doikou15}). Compatibility of the equations (\ref{set1}) leads to the equations of the motion of the system expressed as:
\be
{\partial L_n(t;\lambda) \over \partial t}= V_{n+1}(t;\lambda)\ L_n(t;\lambda) - L_n(t;\lambda)\ V_n(t;\lambda). \label{em1}
\ee

Let us now recall the B\"{a}cklund transformation in this context seen as gauge transformation that leaves the auxiliary linear problem invariant. Indeed, let $M$ be in general an $N\times N$ matrix (Darboux matrix) \cite{sklyanin-back}, such that
\be
\tilde \Psi_n(t;\lambda) = M_n(t;\lambda, \Theta)\ \Psi_n(t;\lambda) \label{darboux}
\ee
where $\Theta$ is an arbitrary constant.

Then the auxiliary problem reads as
\ba
&&\tilde \Psi_{n+1}(t;\lambda) = \tilde L_n(t;\lambda)\ \tilde \Psi_n(t;\lambda)\cr
&& {\partial \tilde \Psi_n(t;\lambda) \over \partial t} = \tilde V_n(t;\lambda)\ \tilde \Psi_n(t;\lambda)
\label{tset1}
\ea
where $\tilde L, \tilde V$ are the gauge transformed matrices. Compatibility of (\ref{set1}), (\ref{tset1}) together  with (\ref{darboux}) lead then to the semi-discrete B\"{a}cklund transformations:
\ba
&& M_{n+1}(t;\lambda,\Theta)\ L_n(t;\lambda) = \tilde L_n(t;\lambda)\ M_{n}(t;\lambda,\Theta) \cr
&& {\partial M_n(t;\lambda,\Theta) \over \partial t} = \tilde V_n(t;\lambda)\ M_{n}(t;\lambda,\Theta) -  M_n(t;\lambda,\Theta)\ V_{n}(t;\lambda).
\label{BT1}
\ea
Notice the similarity of the second B\"{a}cklund equations (\ref{BT1}) with the equations of motion (\ref{em1}). This will be a key point, especially when discussing the issue of defects.

Let us also briefly discuss the fully discrete case, where both space and time are discrete parameters, hence two discrete indices $n,\ m$ are entailed, corresponding to the discrete space and time. This is  the most intricate case, nevertheless we also provide a brief outline here too. The auxiliary linear problem for the discrete Lax pair $L_{nm},\ A_{nm}$ reads as:
\ba
&&\Psi_{n+1m}(\lambda) = L_{nm}(\lambda)\ \Psi_{nm}(\lambda) \cr
&& \Psi_{nm+1}(\lambda) = A_{nm}(\lambda)\ \Psi_{nm}(\lambda).
\label{set2}
\ea
Compatibility of the equations above leads to the discrete equations of motion:
\be
L_{nm+1}(\lambda)\ A_{nm}(\lambda) = A_{n+1m}(\lambda)\ L_{nm}(\lambda). \label{EM2}
\ee
The discrete B\"{a}cklund transformation: consider the discrete Darboux matrix $M_{nm}$ such that
\be
\tilde \Psi_{nm}(\lambda) = M_{nm}(\lambda,\Theta)\ \Psi_{nm}(\lambda)
\ee
provided that the linear auxiliary problem remain invariant under the gauge transformation the fully discrete equations for the B\"{a}cklund transformation arise, expressed as
\ba
&& M_{n+1m}(\lambda,\Theta)\ L_{nm}(\lambda) = \tilde L_{nm}(\lambda)\ M_{nm}(\lambda,\Theta) \cr
&& M_{nm+1}(\lambda,\Theta)\ A_{nm}(\lambda) = \tilde A_{nm}(\lambda)\ M_{nm}(\lambda,\Theta).
\ea
Again the similarity of the latter equations with the discrete equations of motion (\ref{EM2}) is evident. We shall discuss the completely discrete case in full detail elsewhere, especially in the context of time-space ``duality'' along the lines described in \cite{ACDK}.

It is clear that taking the continuum limit in ``time'' we recover the semi-discrete case described previously, i.e. consider
\ba
&& f_{nm}(\lambda) \to f_n(t;\lambda)\cr
&& f_{nm+1}(\lambda) \to f_{n}(t+\delta;\lambda)
\ea
and
\be
A_{nm}(\lambda) \to 1 +\delta V_{n}(t;\lambda)
\ee
where $\delta$ is the discrete ``time'' spacing parameter taken $\delta\to 0$ at the continuum limit.

\subsubsection{The continuous case}
Let us finally discuss the continuous case, which may be also seen as a suitable continuum limit of the case above via ``linearization'': ($L_{nm} \to 1 +\delta U(x,t)$, $A_{nm} \to 1 +\tilde \delta V(x,t)$). The Lax pair in this case $U,\ V$ satisfies (see e.g. \cite{FT})
\ba
&& {\partial \Psi(x,t;\lambda)\over \partial x}  = U(x,t;\lambda)\ \Psi(x,t;\lambda) \label{eq1} \\
&& {\partial \Psi(x,t;\lambda)\over \partial t}  = V(x,t;\lambda)\ \Psi(x,t;\lambda), \label{eq2}
\ea
and the corresponding equations of motion are given by the familiar zero curvature condition
\be
U_t(x,t;\lambda) - V_x(x,t;\lambda) + \Big [U(x,t;\lambda),\ V(x,t;\lambda) \Big] =0,
\ee
where we use the notation: $F_t = {\partial F \over \partial t},\ F_x = {\partial F \over \partial x}$. The Darboux matrix with continuous $(x,\ t)$ parameters transforms the auxiliary function as (see also \cite{wahlq} and \cite{matveev} and references therein)
\be
\tilde \Psi(x, t; \lambda) = M(x, t; \lambda, \Theta)\ \Psi(x,t;\lambda)
\ee
$\Theta$ as in the discrete case is an extra arbitrary parameter; $M$ may be thought of as a local gauge transformation.

Then due to the invariance of the auxiliary linear problem under the transformation one obtains the equations for the B\"{a}cklund transformation:
\ba
&& {\partial M(x,t;\lambda, \Theta) \over \partial x} = \tilde U(x,t;\lambda)\ M(x,t;\lambda,\Theta) - M(x,t;\lambda, \Theta)\ U(x,t;\lambda)\cr
&& {\partial M(x,t;\lambda, \Theta) \over \partial t} = \tilde V(x,t;\lambda)\ M(x,t;\lambda,\Theta) - M(x,t;\lambda, \Theta)\ V(x,t;\lambda). \label{BT2}
\ea
Solving the above set of equations provides the explicit form of the B\"{a}cklund transformation for the system under consideration.

\subsection{Underlying Poisson structure $\&$ space-time dualities}

Having reviewed the auxiliary linear problem and the B\"{a}cklund transformations as gauge transformations of the problem we can now recall the existence of the underlying classical algebras associated to the problem. Let us note that we are going to focus here on ultra-local and non-dynamical classical algebras. And although such generalizations can be in principle considered they will be left for future investigation given that they are outside the scope of the present investigation.

Let us first consider the semi-discrete case. It is well known that in the discrete case one may start from the $L$-matrix of the Lax pair and assuming that $L$ satisfies the quadratic algebra, at equal ``spaces'' (ultra-local algebra):
\be
\Big \{L_{1n}(t;\lambda),\ L_{2n'}(t;\mu)\Big \}_S = \Big [r_{12}(\lambda -\mu),\ L_{1n}(t;\lambda)L_{2n'}(t;\mu) \Big] \delta_{nn'} \label{Fund1}
\ee
the indices $1,\ 2$ are associated to the ``auxiliary'' space, and the $r$-matrix satisfies the classical Yang-Baxter equation. One then can build discrete space monodromies
\be
T_S(\lambda) = L_N(\lambda) \ldots L_1(\lambda),
\ee
$T_S$ yields the charges in involution guaranteeing the integrability of the system under study. The subscript $S$ denotes space-like description (the interested reader is referred to \cite{ACDK} for more details on the ``dual'' description of classical integrable systems).

In the semi-discrete case the $V$-operator provides essentially the time evolution of the system. Indeed, one may consider time-like monodromies
\be
T_{\mathrm T}(n, t_1, t_2;\lambda) = {\cal P} \exp \Big \{ \int_{t_2}^{t_1} V_n(t;\lambda)\ dt \Big \}, ~~~~~t_1 > t_2.
\ee
clearly the subscript ${\mathrm T}$ denotes time-like description. Assuming that $V$ satisfies equal times Poisson structure \cite{ACDK}:
\be
\Big \{ V_{1n}(t;\lambda),\ V_{2n}(t';\mu) \Big \}_{\mathrm T} = \Big [r_{12}(\lambda -\mu),\ V_{1n}(t;\lambda) + V_{2n}(t';\mu) \Big ]\delta(t-t') \label{Fund2}
\ee
$r$ is the same classical $r$-matrix appearing in (\ref{Fund1}).
To conclude in the semi-discrete case the two main assumptions for the Lax pair are given in
(\ref{Fund1}), (\ref{Fund2}).

Similarly, in the fully discrete case monodromies along the discrete time direction can be built assuming that the Lax associated Lax pair satisfies the following set of algebraic relations along the discrete space and time directions:
\ba
&& \Big \{L_{1nm}(\lambda),\ L_{2n'm}(\mu) \Big \}_S = \Big [r_{12}(\lambda - \mu), \ L_{1nm}(\lambda)L_{2n'm}(\mu)\Big ]\delta_{nn'} \cr
&& \Big \{A_{1nm}(\lambda),\ A_{2nm'}(\mu) \Big \}_{\mathrm T} = \Big [r_{12}(\lambda - \mu),\ A_{1nm}(\lambda)A_{2nm'}(\mu)\Big ]\delta_{mm'}.
\ea
The corresponding monodromies are then defined as
\ba
&& T_S(\lambda) = L_{Nm}(\lambda)\ A_{N-1m}(\lambda) \ldots A_{1m}(\lambda)\cr
&& T_{\mathrm T}(\lambda) = A_{nM}(\lambda)\ A_{nM-1}(\lambda) \ldots A_{n1}(\lambda).
\ea
We conclude then that $T_S,\ T_{\mathrm T}$ provide the ``space'' and ``time'' evolution of the system so that a 2D $N \times M$ classical lattice can be then obtained.

Let us finally review the results in the continuum case based on the analysis in \cite{ACDK}. The space and time Poisson structure for the Lax pairs are given by the linear algebraic relations below:
\ba
&& \Big \{U_1(x,t;\lambda),\ U_2(x',t;\mu)\Big \}_S = \Big [r_{12}(\lambda -\mu),\ U_1(x,t;\lambda) + U_2(\mu;x',t) \Big ]\delta(x-x') \label{Fundb1} \\
&& \Big \{V_1(x,t;\lambda),\ V_2(x,t';\mu)\Big \}_T = \Big [r_{12}(\lambda -\mu),\ V_1(x,t;\lambda) + V_2(\mu;x,t'). \Big ] \delta(t-t'). \label{Fundb2}
\ea
Then space and time monodromies may be constructed as solutions of the equations (\ref{eq1}), (\ref{eq2}) respectively:
\ba
&& T_S(x_1, x_2, t;\lambda) = {\cal P} \exp\Big \{ \int_{x_2}^{x_1} U(x,t;\lambda)\ dx\Big \}, ~~~~x_1 > x_2\cr
&& T_{\mathrm T}(x, t_1, t_2;\lambda) = {\cal P} \exp\Big \{ \int_{t_2}^{t_1}  V(x,t;\lambda)\ dt \Big \}, ~~~~~t_1>t_2.
\ea

Given that space and time monodromies in both discrete and continuous integrable models satisfy quadratic algebras we also have:
\be
\Big \{T_{1}(\lambda),\ T_{2}(\mu) \Big \}_{S, T} = \Big [ r_{12}(\lambda -\mu),\ T_{1}(\lambda)\ T_2(\mu) \Big ] \label{basic2}
\ee
we conclude that the time and space like Poisson involution is guaranteed, i.e.
\ba
&& \Big \{ tr (T_{S, {\mathrm T}}(\lambda)),\ tr(T_{S, {\mathrm T}}(\mu)) \Big \}_{S,T} = 0, \cr
&& x \in  [-{\mathrm L},\ {\mathrm L}], ~~~ t \in [-{\mathrm T},\ {\mathrm T}]~~ {\mathrm L},\ {\mathrm T} \to \infty
\ea
$tr (T_{{\mathrm T},S})$ provide the charges in involution, i.e. Liouville integrability is ensured.

Moreover, as is well known starting from the $L$ or $U$ matrix and assuming that they satisfy the algebraic constraints described above one may obtain the generating function of the hierarchy $V$-operators associated to each one of the integrals of motion \cite{sts}, indeed the relevant expressions of the time components of the Lax pair in the semi-discrete case are given by \cite{sts}
\ba
&& V_{2n}(t;\lambda,\mu) = t_S^{-1}(\lambda)\ tr\Big [T_{1S}(N,n;\lambda)\ r_{12}(\lambda -\mu)\ T_{1S}(n-1,1;\lambda) \Big ], \cr
&& t_S(\lambda) = tr (T_S(\lambda))\label{V1}
\ea
where the following notation is introduced
\be
T(n,m;\lambda) = L_n(\lambda) \ldots L_m(\lambda), ~~~~n>m.
\ee
In the continuum case the generating function of the $V$-operators are derived from \cite{sts, FT}
\be
V_{2}(x,t;\lambda,\mu) = t_S^{-1}(\lambda)\ tr\Big [T_{1S}({\mathrm L},x,t;\lambda)\ r_{12}(\lambda -\mu)\ T_{1S}(x,-{\mathrm L}, t;\lambda) \Big ]. \label{V2}
\ee
considering $x \in [-{\mathrm L},\ {\mathrm L}]$.
Expansion of the $V$-operator in powers of $\lambda$ provides the time components Lax pairs associated to each integral of motion.

Similarly, one can also start from $V$ and assuming that satisfies the ``time-like'' algebraic constraints discussed earlier in the text obtain the relevant $U$ operators as was shown in \cite{ACDK}, this is the so-called ``dual'' description. The relevant $U$ expressions in this case then are provided by \cite{ACDK}
\ba
&& U_{2}(x,t;\lambda ,\mu) = t_{\mathrm T}^{-1}(\lambda)\ tr\Big [T_{1{\mathrm T}}(x,{\mathrm T}, t;\lambda)\ r_{12}(\lambda -\mu)\ T_{1{\mathrm T}}(x,t,-{\mathrm T};\lambda) \Big ],\cr
&& t_{\mathrm T}(\lambda) = tr (T_{\mathrm T}(\lambda)). \label{U1}
\ea
$t \in [-{\mathrm T},\ {\mathrm T}]$. In \cite{ACDK} only the continuous case was considered, however in principle a similar construction should be possible in the context of (semi)-discrete integrable models. We shall not comment further on the point here due to certain subtleties, which will be hopefully addressed in a separate publication.

\subsubsection*{A note on the algebraic content of B\"{a}cklund transformations}

The algebraic content of the B\"{a}cklund transformations will be presented in this section given also the similarity with defect matrices, which will be discussed in the subsequent section in more detail. The defect matrices satisfy certain algebraic constraints consistent with the integrability of the physical system under consideration.
Let us now focus on the algebraic content of the Darboux matrices $M$ in both continuous and discrete cases. One can show in a straightforward manner exploiting the fact that the auxiliary function satisfies a classical version of the vertex algebra \cite{doikou15}, that the Darboux matrix $M$ also satisfies certain classical quadratic algebras. In \cite{doikou15} the discussion was restricted in the continuous case yielding for the auxiliary function the following:
\be
\Big \{\Psi_1(x,t;\lambda),\ \Psi_2(x,t;\mu)\Big \} = \mathfrak{r}_{12}(\lambda -\mu)\ \Psi_1(x,t; \lambda)\ \Psi_2(x,t;\mu), ~~~{\mathfrak r}(\lambda) = r(\lambda) - f(\lambda),
\ee
where $r$ is a solution of the classical Yang-Baxter equation, and $f$ is a function depending on the choice of the $r$-matrix.

Using similar arguments as in \cite{doikou15} one can show for the semi discrete auxiliary function:
\be
\Big \{\Psi_{1n}(t;\lambda),\ \Psi_{2n}(t;\mu)\Big \} = \mathfrak{r}_{12}(\lambda -\mu)\ \Psi_{1n}(t; \lambda)\ \Psi_{2n}(t;\mu).
\ee
Assume that the same Poisson structure should be satisfied by the gauge transformed auxiliary function $\tilde \Psi = M\ \Psi$ in both the continuous and the discrete case, and also
\ba
&& \Big\{M_1(x;\lambda),\ \Psi_2(x;\mu) \Big \} =  0, ~~~~~\mbox{continuous}\cr
&& \Big\{M_{1n}(\lambda),\ \Psi_{2n}(\mu) \Big \} =  0, ~~~~~\mbox{discrete}.
\ea
Then due to consistency requirements it is straightforward to show that the Darboux matrix $M$ satisfies the following quadratic Poisson algebras:
\ba
&& \Big \{M_1(x; \lambda),\ M_2(x, \mu) \Big \} =  \Big [r_{12}(\lambda-\mu),\ M_1(x,\lambda)\ M_2(x,\mu)\Big] ~~~~\mbox{continuous}\cr
&& \Big \{M_{1n}(\lambda),\ M_{2n}(\mu) \Big \} = \Big [r_{12}(\lambda-\mu),\ M_{1n}(\lambda)\ M_{2n}(\lambda)\Big]~~~~~\mbox{discrete}.
\ea
From the duality picture described in \cite{ACDK} the time-like monodromies provide for the auxiliary function the time like Poisson structure as well.

As will be clear in what follows similarly to the case of defects classification of representations of the quadratic classical algebras will be the first step into identifying the suitable Darboux matrices, then solution of the B\"{a}cklund relations will lead to the full identification of the B\"{a}cklund transformation. The auxiliary linear problem remains to be understood at the quantum level as well. In this context it would be very interesting to explain how local Darboux matrices may lead to the quantum analogue of the generating function of the B\"{a}cklund transformation i.e. the $Q$ operator along the lines described in \cite{sklyanin-back}. This issue is mostly relevant especially in the fully discrete case where the $Q$ operator may be thought of as the discrete time evolution operator\footnote{I am grateful to C. Korff for useful discussions and insightful comments on this subject}.

\section{Implementing defects}

We come now to our main objective which is the interpretation of local defects as quasi B\"{a}cklund transformations using as fundamental building blocks the associated auxiliary linear problem as well as the underlying classical algebra. The similarities between the equations of motion associated to the defect degrees of freedom and the space and time parts of the B\"{a}cklund transformation as discussed in the previous section will be manifest in what follows.  Let us first recall the formulation of integrable defects in both discrete and continuous theories.

\subsection{The discrete case}

We focus on the semi discrete case (discrete space, continuous time) and assume that we insert a defect at the $n^{th}$ cite of the classical system. Then the associated linear auxiliary problem is modified as follows (see also \cite{avan-doikou}):
\ba
&& \Psi_{j+1}(t; \lambda)= L_j(t; \lambda)\ \Psi_j(t; \lambda) \cr
&& {\partial \Psi_j(t;\lambda)\over \partial t} = V_j(t;\lambda)\ \Psi_j(t; \lambda), ~~~~ j\neq n \label{lineardefect}
\ea
and on the defect point in particular the problem is formulated as:
\ba
&&\Psi_{n+1}(t;\lambda) = \tilde L_n(t;\lambda)\ \Psi_n(t;\lambda) \cr
&& {\partial \Psi_n \over \partial t} = V_n(t;\lambda)\Psi_n(t; \lambda)\label{lineradefectb}
\ea
where $\tilde L_n$ is the defect matrix. Compatibility conditions of equations (\ref{lineradefectb}), provide the time evolution of the degrees of freedom associated to the defect:
\be
{\partial \tilde L_n(t;\lambda) \over \partial t} = V_{n+1}(t;\lambda)\ \tilde L_n(t;\lambda) - \tilde L_n(t; \lambda)\ V_n(t;\lambda).
\ee
The latter resembles structurally the time component of the B\"{a}cklund transformation (\ref{BT1}). One may see the defect as a {\it quasi} B\"{a}cklund transformation dividing the one dimensional space into two distinct areas with different solutions of the same or different integrable differential equations provided that all the Lax-pairs involved satisfy classical algebras that entail the same $r$-matrix. The suitable set of algebraic relations associated to this setting will be explicitly formulated.

The other part of the B\"{a}cklund transformation, i.e. the ``space'' component
can be then obtained via the ``dual'' picture as described in \cite{ACDK}, in which one considers the existence of a defect along the time axis at $t=t_0$. Let $\tilde A_j(t_0)$ be the defect matrix along the time direction, then the auxiliary linear problem is formulated as follows:
\ba
&& \Psi^{\pm}_{j+1}(t;\lambda) =L^{\pm}_j(t;\lambda)\ \Psi_j^{\pm}(t;\lambda)\cr
&& {\partial \Psi_j^{\pm}(t;\lambda) \over \partial t} = V_j^{\pm}(t;\lambda)\ \Psi_j^{\pm}(t;\lambda), ~~~~t\neq t_0. \label{linearbb}
\ea
In general, we introduce the notation for any function of time $F$: $F^+$, for $t> t_0$ and $F^-,$ for $t<t_0$. On the defect point we have:
\ba
\Psi^+_j(t_0; \lambda) = \tilde A_j(t_0;\lambda)\ \Psi^-_j(t_0;\lambda)
\ea
Compatibility of the latter equation with the first of the equations (\ref{linearbb}) leads to
\be
L_j^+(t_0;\lambda)\  \tilde A_j(t_0; \lambda) =  \tilde A_{j+1}(t_0)\ L_j^-(t_0;\lambda),
\ee
which is similar to the discrete space part of the B\"{a}cklund transformation equations in the semi discrete case (\ref{BT1}).
The similarity is striking one has to just make the following identifications: $\tilde A_j \to M_j$. It is however important to note that unlike the usual B\"{a}cklund transformation relations, where the time and space parts are simultaneously satisfied, in the case of defects the time and space relations (local equations of motion) derived above are separately satisfied except on the particular point in 2D space-time i.e. $(t_0,\ x_0)$ .

Let us also briefly consider the fully discrete case. We introduce the defect along the discrete space direction via the defect matrix $\tilde L_{n_0m}$. Then on the defect point the auxiliary problem reads as
\ba
&& \Psi_{n_0+1m}(\lambda) = \tilde L_{n_0m}(\lambda)\ \Psi_{n_0m}(\lambda)\cr
&& \Psi_{n_0 m+1}(\lambda) = A_{n_0 m}(\lambda)\ \Psi_{n_0m}(\lambda),
\ea
and the equations of motion on the defect point are then:
\be
\tilde L_{n_0 m+1}(\lambda)\ A_{n_0m}(\lambda) = A_{n_0+1 m}(\lambda)\ \tilde L_{n_0m}(\lambda).
\ee
Similarly if we choose to introduce a defect along the discrete time direction via the defect matrix $\tilde A_{nm_0}$ then the auxiliary problem becomes:
\ba
&& \Psi_{n+1 m_0}(\lambda) = L_{nm_0}(\lambda)\ \Psi_{nm_0}(\lambda) \cr
&& \Psi_{nm_0+1}(\lambda) = \tilde A_{nm_0}(\lambda)\ \Psi_{nm_0}(\lambda)
\ea
and the corresponding equations of motion are then given as:
\be
L_{nm_0+1}(\lambda)\ \tilde A_{nm_0}(\lambda) = \tilde A_{n+1 m_0}(\lambda)\ L_{nm_0}(\lambda).
\ee
Again this part of the equations of motion is very similar to the one of the parts of the fully discrete B\"{a}cklund transformation. The defect in any case may be thought of as already mentioned as an object that divides the one dimensional discrete space (or time) into two regions with distinct solutions of the same or even different difference equations.

\subsection{The continuous case}

Let us complete the discussion about our setting with the continuous case in the presence of point-like defects. We start with the space like description of the problem and we implement a defect along the $x$ axis at the point $x_0$. The auxiliary problem away from the defect point (see also e.g. \cite{caudrelier, avan-doikou})
\ba
&& {\partial \Psi^{\pm}(x,t;\lambda) \over \partial x} = U^{\pm}(x,t;\lambda)\ \Psi^{\pm}(x,t;\lambda) \cr
&& {\partial \Psi^{\pm}(x,t;\lambda) \over \partial t} = V^{\pm}(x,t;\lambda)\ \Psi^{\pm}(x,t;\lambda), ~~~~x\neq x_0
\ea
we introduce the notation: for any function of $x,\ t $ we have $F^+$ for $x > x_0$ and $F^-$ for $x< x_0$.
On the defect point the problem is formulated as:
\ba
\Psi^+(x_0,t;\lambda) = \tilde L(x_0, t;\lambda,\Theta)\ \Psi^{-}(x_0, t;\lambda)
\ea
$\Theta$ an arbitrary constant. Compatibility condition between the latter equations on the defect point provides:
\be
{\partial \tilde L(x_0, t;\lambda,\Theta) \over \partial t} = \tilde V^+(x_0, t;\lambda)\ \tilde L(x_0, t;\lambda,\Theta)- \tilde L(x_0, t;\lambda, \Theta)\ \tilde V^-(x_0, t;\lambda) \label{qBT2t}
\ee
where $\tilde V^{\pm}$ are the Lax pair time components computed around the defect point (see more details on this matter in \cite{avan-doikou}). Note that around the defect point necessary analyticity conditions were imposed (see \cite{avan-doikou} for more details)
\be
\tilde V^{\pm}(x_0) \to V^{\pm}(x_0^{\pm})
\ee
leading to certain ``jump'' like conditions between the fields and their derivatives. Also, (\ref{qBT2t}) resembles the time part of the B\"{a}cklund transformation in (\ref{BT2}).

To obtain the $x$-part of the B\"{a}cklund transformation we shall consider the ``dual'' time-like description.
Indeed, from the time-like perspective we impose a defect along the $t$ axis at $t =t_0$, then on the defect point we have
\be
\psi^{\pm}(x,t_0;\lambda) = \tilde A(x, t_0; \lambda, \Theta)\ \Psi^{-}(x, t_0; \lambda).
\ee
Via the compatibility of the latter equations with the bulk auxiliary linear problem around the defect point provide the $x$ part of the B\"{a}cklund transformation (\ref{BT2}):
\be
{\partial \tilde A(x, t_0;\lambda, \Theta) \over \partial x} = U^+(x,t_0; \lambda)\ \tilde A(x, t_0; \lambda, \Theta) - \tilde A(x,t_0; \lambda, \Theta)\ U^-(x, t_0;\lambda). \label{qBT2x}
\ee
Algebraic conditions on the Lax pairs and on the defect matrix $\tilde L,\ \tilde A$ along the $x$ and $t$ line guaranteeing a la Liouville integrability will be discussed in the next subsection for both discrete and continuous models. Let us note that one may choose to consider the matrices $\tilde L,\ \tilde A$ to coincide given that these are structurally similar objects. This is paricularly relevant if one considers the $x$ and $t$ evolution equations on the $(x_0,\ t_0)$ point, where both equations are satisfied simultaneously exactly as in the familiar B\"{a}cklund transformation setting. However if the defect is considered to be ``fixed'' then no time or space evolution occurs for the defect degrees of freedom. This happens only if we allow the defect to ``move'' in space and time.

\subsection{Underlying Poisson structure}

Let us demonstrate the algebraic content of the defect matrices in both space and time directions. The main conditions in addition to the ``bulk'' algebraic relations (\ref{Fund1}), in the semi-discrete in the space-like and time-like description respectively are given as ($\tilde L,\ \tilde A$ are also $\Theta$ dependent, but the dependence is omitted below for brevity)
\ba
&&\Big \{ \tilde L_{1n}(t;\lambda),\ \tilde L_{2n}(t;\mu)\Big \} = \Big [ r_{12}(\lambda -\mu),\ \tilde L_{1n}(t;\lambda)\tilde L_{2n}(t, \mu) \Big ] \cr
&& \Big \{ \tilde A_{1j}(t_0; \lambda),\ \tilde A_{2j}(t_0;\mu)\Big \} = \Big [ r_{12}(\lambda -\mu),\ \tilde A_{1j}(t_0;\lambda)\tilde A_{2j}(t_0; \mu) \Big ]. \label{dFund1}
\ea
The modified space and time monodromies for the semi-discrete case, in the presence of defects
\ba
&& T_S(\lambda) = L_{N}(\lambda)\ldots \tilde L_n(\lambda) \ldots L_1(\lambda)\cr
&& T_{\mathrm T}(\lambda) = {\cal P} \exp\Big \{\int_{t_0^+}^{{\mathrm T}} V^{+}(x,t;\lambda) dt \Big \}\
\tilde A(x, t_0; \lambda)\ {\cal P} \exp\Big \{\int_{- {\mathrm T}}^{t_0^-} V^{-}(x,t;\lambda) dt \Big \}.
\ea

Similarly, in the fully discrete case the algebraic constraints for $\tilde L$ and $\tilde A$ are given as
\ba
&&\Big \{ \tilde L_{1n_0m}(\lambda),\ \tilde L_{2n_0m}(\mu)\Big \}_S = \Big [ r_{12}(\lambda -\mu),\ \tilde L_{1n_0m}(\lambda)\tilde L_{2n_0m}( \mu) \Big ] \cr
&& \Big \{ \tilde A_{1nm_0}(\lambda),\ \tilde A_{2nm_0}(\mu)\Big \}_{\mathrm T} = \Big [ r_{12}(\lambda -\mu),\ \tilde A_{1nm_0}(\lambda)\tilde A_{2nm_0}(\mu) \Big ]
\ea
with discrete space and time modified monodromies defined as
\ba
&& T_S(\lambda) = L_N(\lambda) \ldots \tilde L_n(\lambda) \ldots L_1(\lambda)\cr
&& T_{\mathrm T}(\lambda) = A_M(\lambda) \ldots \tilde A_m(\lambda) \ldots A_1(\lambda).
\ea

Finally, in the continuum case the algebraic relations that hold for the $\tilde L$ and $\tilde A$ matrices on a specific space and time point respectively:
\ba
&&\Big \{ \tilde L_{1}(x_0, t;\mu),\ \tilde L_{2}(x_0,nt;\lambda)\Big \} = \Big [ r_{12}(\lambda -\mu),\ \tilde L_{1}(x_0,t;\lambda)\tilde L_{2}(x_0, t, \mu) \Big ] \cr
&& \Big \{ \tilde A_{1}(x, t_0; \lambda),\ \tilde A_{2}(x, t_0;\mu)\Big \} = \Big [ r_{12}(\lambda -\mu),\ \tilde A_{1}(x, t_0;\lambda)\tilde A_{j}(x, t_0; \mu) \Big ], \label{dFund2}
\ea
and the corresponding modified monodromies are defined as
\ba
&& T_S(\lambda) = {\cal P} \exp\Big \{\int_{x_0^+}^{{\mathrm L}} U^{+}(x,t;\lambda) dx \Big \}\
\tilde L(x_0, t; \lambda)\ {\cal P} \exp\Big \{\int_{- {\mathrm L}}^{x_0^-} U^{-}(x,t;\lambda) dx \Big \} \cr
&& T_{\mathrm T}(\lambda) = {\cal P} \exp\Big \{\int_{t_0^+}^{{\mathrm T}} V^{+}(x,t;\lambda) dt \Big \}\
\tilde A(x, t_0; \lambda)\ {\cal P} \exp\Big \{\int_{- {\mathrm T}}^{t_0^-} V^{-}(x,t;\lambda) dt \Big \}.
\ea
Given the above algebraic conditions one may show that the trace of the modified monodromies provide charges in involution in the both time or space direction,
\ba
\Big \{tr(T_{S,{\mathrm T}}(\lambda),\ tr(T_{S,{\mathrm T}}(\mu)) \Big\}_{S, {\mathrm T}} = 0.
\ea

As discussed in the bulk case in the previous section starting from the $L$ or $U$ matrix and assuming that they satisfy certain algebraic constraints one may obtain the various $V$ operators associated to each one of the integrals of motion \cite{sts}. Similarly in the presence of point like defects one can construct the time component of the Lax pair given the $L$ or $U$ and $\tilde L$ matrices and the associated algebraic constraints. The relevant expressions of the time components of the Lax pair in the semi-discrete case in the presence of defects are given by \cite{avan-doikou}
\ba
&& V_{2n}(t;\lambda, \mu) = t_S^{-1}(\lambda)\ tr\Big [T_{1S}(N,n+1;\lambda)\ \tilde L_{1n}(\lambda)\ r_{12}(\lambda -\mu)\ T_{1S}(n-1,1;\lambda) \Big ] \cr
&&  V_{2n+1}(t;\lambda, \mu) = t_S^{-1}(\lambda)\ tr\Big [T_{1S}(N,n+1;\lambda)\ r_{12}(\lambda -\mu)\ \tilde L_{1n}(\lambda)\ T_{1S}(n-1,1;\lambda) \Big ]. \nonumber\\
\ea

In the continuum case the generating function of the $V$-operators computed on the defect point from the left and from the right are given as  \cite{avan-doikou}
\ba
&&  \tilde V^-_{2}(x_0,t;\lambda) = t^{-1}(\lambda)\ tr\Big [T_{1S}(L,x_0,t;\lambda)\ L_1(x_0, t ;\lambda)\  r_{12}(\lambda -\mu)\ \tilde  T_{1S}(x_0,-L, t;\lambda) \Big ] \cr
&& \tilde V^+_{2}(x_0,t;\lambda) = t^{-1}(\lambda)\ tr\Big [T_{1S}(L, x_0,t;\lambda)\ r_{12}(\lambda -\mu)\ \tilde L_1(x_0, t ;\lambda)\ T_{1S}(x_0,-L, t;\lambda) \Big ] \nonumber\\
\ea
considering $x_0 \in [-{\mathrm L},\ {\mathrm L}]$. $\tilde V^{\pm}$ as we have already commented before are the matrices computed on the defect point from the left and from the right. The bulk quantities are given by the usual expressions (\ref{V2}).

Similarly, one can consider the time-like approach start from $V, \tilde A$ and assuming that they satisfy time-like algebraic constraints
obtain the relevant $U$ operators. The relevant $U$ expressions are then given as around the defect point
\ba
&& \tilde  U^-_{2}(x,t_0;\lambda) = t_{\mathrm T}^{-1}(\lambda)\ tr\Big [T_{1{\mathrm T}}(x,{\mathrm T}, t_0;\lambda)\ \tilde A_1(x, t_0, \lambda)\ r_{12}(\lambda -\mu)\ T_{1{\mathrm T}}(x,t_0,-{\mathrm T};\lambda) \Big ] \cr
&& \tilde U^+_{2}(x,t_0;\lambda) = t_{\mathrm T}^{-1}(\lambda)\ tr\Big [T_{1{\mathrm T}}(x,{\mathrm T}, t_0;\lambda)\ r_{12}(\lambda -\mu)\ \tilde A_1(x, t_0, \lambda)\ T_{1{\mathrm T}}(x,t_0,-{\mathrm T};\lambda) \Big ] \cr &&
\ea
$t_0 \in [-{\mathrm T},\ {\mathrm T}]$. Similarly to the space-like description analyticity conditions around the defect point:
\be
\tilde U^{\pm}(t_0) \to U^{\pm}(t_0^{\pm})
\ee
lead to ``time jump'' conditions on the fields and their time derivatives. The bulk quantities coincide with the ones given in (\ref{U1}).

Here as in \cite{ACDK} we focus only on expressions regarding the continuous case. Similar expressions may be in principle identified in the discrete case, but this is left as a separate investigation given that it involves certain intricate issues.

\section{Two examples}
In order to demonstrate the formulation of the local defects as quasi B\"{a}cklund transformations
we provide below two specific examples of discrete and continuous cases, where defects are treated as (quasi) B\"{a}cklund transformations. In particular, we shall focus on two prototype integrable models, the Toda chain and the sine-Gordon model.

\subsection{The Toda chain}
Consider the Toda chain in the presence of a space like defect.
We insert the local defect represented by the defect matrix $\tilde L$ at the $n^{th}$ cite of the one dimensional system. The bulk Lax pair of the model is given by:
\ba
L_j(\lambda)=
\begin{pmatrix}
 \lambda - p_j  & e ^{q_j} \cr
 -e^{-q_j} & 0
\end{pmatrix},~~~~~j \neq n. \label{pair}
\ea
The space Poisson commutator for the dynamical quantities $q_i,\ p_i$ emanate from the algebraic relations (\ref{Fund1}) with the $r$-matrix being the Yangian \cite{yang},
\be
r(\lambda) = {\mathrm P \over \lambda}, ~~~{\mathrm P} ~~~~~\mbox{the permutation operator}.
\ee
then
\be
\Big \{q_i,\ p_j \Big \}_S = \delta_{ij}.
\ee
We may implement two types of defects, i.e. type I given by the matrix
\ba
\tilde L^{(I)}_{n}(\lambda)  = \begin{pmatrix}
 \lambda -\Theta + {\mathbb N}_n  & x_n\cr
 -X_n & 0
\end{pmatrix},
\ea
and type II with with the $L$ matrix given as
\ba
\tilde L^{(II)}_{n}(\lambda)  = \begin{pmatrix}
 \lambda -\Theta + \alpha_n  & \beta_n\cr
 \gamma_n & \lambda -\Theta - \alpha_n
\end{pmatrix}.
\ea
Given that $\tilde L^{(I, II)}$ satisfy the quadratic algebra (\ref{dFund1}) with the Yangian $r$-matrix
then the Poisson relations for the dynamical degrees of freedom are then given via (\ref{dFund1}) as:
\be
\Big \{x_n,\ X_n \Big \} =1.
\ee
and for the type II defect:
\ba
&& \Big \{\alpha_n,\ \beta_n \Big\} = \beta_n \cr
&& \Big\{\alpha_n\, \gamma_n \Big \} = -\gamma_n\cr
&& \Big\{\beta_n\, \gamma_n \Big \} = 2\alpha_n.
\ea
Expanding the $\log tr (T_S(\lambda))$  in the presence of defect in powers of $\lambda^{-1}$
we obtain the local integrals of motion. Note that type I defects provide rather trivial results, especially in the context of the quasi B\"{a}cklund transformations, therefore we shall focus henceforth on type II results. In particular, the momentum and Hamiltonian for  type II defects are given below (we consider henceforth $\Theta =0$ for simplicity, but without really losing generality)
\ba
&& P = -\sum_{j \neq n} p_j + \alpha_n \cr
&& H = -{1\over 2} \sum_{j \neq j} p_j^2 - \sum_{j \neq n, n-1} e^{q_{j+1} - q_j} - e^{q_{n+1} - q_{n-1}} -\beta_n e^{-q_{n-1}} - \gamma_n e^{q_{n+1}} - {{\alpha_n}^2\over 2}.
\ea
From the expansion of the expression providing the time component of the Lax pair (\ref{pair}) we identify around the defect point:
\ba
&& \tilde V_n = \begin{pmatrix}
 \lambda  & e^{q_{n+1}} + \beta_n\cr
 -e^{-q_{n-1}} & 0
\end{pmatrix} \cr
&& \tilde V_{n+1} =  \begin{pmatrix}
 \lambda  & e^{q_{n+1}}\cr
 \gamma_n-e^{-q_{n-1}} & 0
\end{pmatrix}.
\ea
It is clear that the latter expressions are slightly ``deformed'' compared to the bulk expressions away from the defect (see relevant discussion in \cite{avan-doikou} in the context of the discrete NLS model). The bulk $V$-matrix is given in as
\ba
V_j(\lambda)=
\begin{pmatrix}
 \lambda  & e ^{q_j} \cr
 -e^{-q_{j-1}} & 0
\end{pmatrix} ~~~~~j \neq n,\ n+1. \label{pair2}
\ea

Let us now consider the equations of motion related to the defect degrees of freedom i.e. the time component of the semi-discrete quasi B\"{a}cklund transformation:
\be
{\partial \tilde L_n \over \partial t} = \tilde V_{n+1}\ \tilde L_n(\lambda) - \tilde L_n(\lambda)\ \tilde V_n(\lambda).
\ee
Solving the latter equation provides the following set of relations for the time evolution of the degrees of freedom of the defect:
\ba
&&  \dot \alpha_n  = e^{q_{n+1}}\ \gamma_n + e^{-q_{n-1}}\ \beta_n \cr
&& \dot \beta_n   = -2 \alpha_n e^{q_{n+1}} - \alpha_n \beta_n \cr
&&  \dot  \gamma_n   = -2 \alpha_n e^{-q_{n-1}} +\alpha_n \gamma_n
\ea
the ``dot'' denotes differentiation with respect to time. It is also relevant to consider the equations of motion around the defect point at $n\pm 1$ given that these are slightly modified compared to the bulk ones due to the presence of the ``deformed'' $\tilde V_n,\ \tilde V_{n+1}$, indeed
\ba
&& {\partial L_{n+1}(\lambda) \over \partial t} = V_{n+2}(\lambda)\ L_{n+1}(\lambda) - L_{n+1}(\lambda)\ \tilde V_{n+1}(\lambda) \cr
&& {\partial L_{n-1}(\lambda) \over \partial t} = \tilde V_{n}(\lambda)\ L_{n-1}(\lambda) - L_{n-1}(\lambda)\  V_{n-1}(\lambda),
\ea
which lead respectively to the following extra conditions among the relevant fields and the defect degrees of freedom
\ba
&&  \dot q_{n+1}  = p_{n+1} \cr
&& \dot p_{n+1}   = e^{q_{n+2} -q_{n+1}} - e^{q_{n+1}}(\gamma_n - e^{-q_{n-1}})
\ea
and
\ba
&& \dot q_{n-1}  = p_{n-1}  \cr
&& \dot  p_{n-1}  = -e^{q_{n-1} -q_{n-2}} + e^{-q_{n-1}}(\beta_n + e^{q_{n+1}})
\ea
These equations may be thought ``deformed equations'' of motion around the defect point. Indeed, compare the latter equations with the familiar bulk equations for $j \neq n,\ n\pm 1$
\ba
&& \dot q_j = p_j \cr
&& \dot p_j = e^{q_{j+1}- q_j} - e^{q_j - q_{j-1}}.
\ea
Naturally, the main objective is to solve the equations around the defect point and understand the time evolution and thus the behavior of the defect degrees of freedom. More precisely, we can consider situations with zero, one or two discrete solitons and see how their presence affects the time evolution of the defect degrees of freedom. The detailed analysis of the latter equations will be discussed in detail elsewhere. It is worth noting that as opposed to the continuous case no analyticity conditions around the defect point are considered, thus one requires a ``cluster'' of equations of motion around the defect point in order to examine the time evolution of the defect.

Note also that the space part of the discrete B\"{a}cklund transformation can be obtained via the dual description, however this requires the identification of $L, \tilde L$ starting from $V$. This however is a subtle issue, which will be separately addressed in a forthcoming publication.

\subsection{The sine-Gordon model}

The sine-Gordon model in the presence of local defects has been extensively studied (see e.g. \cite{corrigan-atft, corrigan-qu1, avan-doikou, caudrelier}) during the last decade or so. Time evolution equations were derived for the degrees of freedom of the defect for type II defects. The connection of these equations with the sewing conditions derived in \cite{avan-doikou} as analyticity conditions was apparent, and quite surprising. This will be particularly evident in the explicit computations below when dealing with the time part of the B\"{a}cklund transformation (see e.g. \cite{wahlq, matveev}.

Let us first introduce the Lax pair for the sine-Gordon model:
\ba
&& U(x,t;\lambda) = {1\over 2} \begin{pmatrix}
  -W_t &\sinh(\lambda +W)\cr
  \sinh(\lambda - W)& W_t
\end{pmatrix}, \cr
&&V(x,t;\lambda) = {1\over 2} \begin{pmatrix}
 -W_x  & \cosh(\lambda+ W)\cr
 \cosh(\lambda-W) & W_x
\end{pmatrix}
\ea
where $W = {i\beta \over 2}\phi$, $\phi$ is the sine-Gordon field.
As in the case of the Toda chain, one may consider two types of defects, type I associated to the deformations of the harmonic oscillator algebra and type II associated to a suitable deformations of the $\mathfrak{sl}_2$ algebra. The corresponding defect matrices are given below for the two types of defects:
\ba
\tilde L^{(I)}_{n}(\lambda) = \begin{pmatrix}
 X  & e^{\lambda}Z^{-1}\cr
 e^{\lambda}Z & X^{-1}
\end{pmatrix},
\ea
and the type II defect matrix
\ba
\tilde L^{(II)}_{n}(\lambda) = \begin{pmatrix}
 e^{\lambda}{\cal V} - e^{-\lambda} {\cal V}^{-1} &\bar a \cr
 a & e^{\lambda}{\cal V}^{-1} - e^{-\lambda} {\cal V}
\end{pmatrix}.
\ea
We shall basically focus hereafter on type II defects (see also \cite{corrigan-fused, avan-doikou}) given that type I defect gives rise to the familiar B\"{a}cklund transformation for the sine-Gordon model, and the degrees of freedom of the defect somehow ``disappear'', given that they are eventually exclusively expressed in terms of the left/right bulk fields. It is worth pointing out that the type II defect matrix in this particular context may be expressed as the product of two type I matrices (see e.g. \cite{corrigan-fused} and references therein). This of course will have certain implications when solving the space and/or time part of the B\"{a}cklund transformation, especially given the fact that products of fundamental B\"{a}cklund transformations provide multi-solitonic solutions. This is a significant issue that merits further investigation.

Both types I and II defect matrices satisfy the quadratic algebra (\ref{dFund2}), with the trigonometric $r$-matrix (see \cite{FT} and references therein):
\be
r(\lambda) = {\beta^2 \over 8 \sinh(\lambda)}\begin{pmatrix}
 {\sigma^z +1 \over 2}\cosh(\lambda)  & \sigma^-\cr
 \sigma^+ & {-\sigma^z +1 \over 2}\cosh(\lambda)
\end{pmatrix}.
\ee
One may then deduce the algebraic relations for the degrees of freedom of the defect:
\ba
&& \{ a,\ {\cal V}\}= {\beta^2 \over 8}{\cal V} a\cr
&&  \{ \bar a,\ {\cal V}\}= -{\beta^2 \over 8} {\cal V} \bar a\cr
&&  \{ a,\ \bar a\}= -{\beta^2\over 4} ({\cal V}^2 - {\cal V}^{-2})
\ea
the latter may be thought of as a deformation of the classical $\mathfrak{sl}_2$ algebra.

We shall solve now both B\"{a}cklund transformation relations for the type II defect. Let us first consider the time component of the relations (\ref{qBT2t}) based on the space-like description of the model. Solving (\ref{qBT2t}) we obtain the following set of relations describing the time evolution of the defect degrees of freedom (see also \cite{avan-doikou} for a similar set of relations)
\ba
&& 2 \bar a_t = - \bar a(W^+_x + W^-_x) - 2 \cosh({W^+ + W^- \over 2} - \Theta)\ \sinh (W^+ - W^-) \cr
&& 2 a_t = a (W^+_x + W^-_x) + 2 \cosh({W^+ + W^- \over 2} + \Theta)\ \sinh (W^+ - W^-).
\ea
Also, from the diagonal entries of (\ref{qBT2t})
\ba
&& 2 (\ln {\cal V})_t = W_x^- -W_x^+  +{e^{-\Theta}\over 2}(a e^{W^+ + W^- \over 2} - \bar a e^{-W^+ - W^- \over 2})\cr
&& 2 (\ln {\cal V})_t = W_x^+ - W_x^- +{e^{\Theta}\over 2}(a e^{-W^+ - W^- \over 2} - \bar a e^{W^+ + W^- \over 2}).
\ea
Compatibility conditions of the latter equations lead to the following equation
\be
W_x^+ - W_x^- = {a\over 2} \sinh({W^+ + W^- \over 2} -\Theta)+ {\bar a\over 2} \sinh({W^+ + W^- \over 2} +\Theta).
\ee
The relations above agree with the respective relations derived in \cite{avan-doikou}. It is worth noting that part of these relations arise as analyticity conditions on the time components of the Lax pair, a fact that provides a very strong consistency check of the methodology introduced in \cite{avan-doikou}. Moreover, in \cite{avan-doikou} the time part of the B\"{a}cklund transformation in both discrete and continuous models is basically derived and thought of as providing the local equations of motion on the defect point. As in the discrete case it would be informative to solve the equations using zero, one or two soliton solutions and obtain the respective time evolution of the dynamical degrees of freedom of the defect.

Here, we also provide the dual picture along the lines described in \cite{ACDK}, giving rise to the $x$ component of the B\"{a}cklund transformation (\ref{qBT2x})
\ba
&& 2\bar a_x = - \bar a(W^+_t + W^-_t) - 2 \sinh({W^+ + W^- \over 2} - \Theta)\ \sinh (W^+ - W^-) \cr
&& 2 a_x = a (W^+_t + W^-_t) - 2 \sinh({W^+ + W^- \over 2} + \Theta)\ \sinh (W^+ - W^-).
\ea
Also, from the diagonal entries of the matrices in (\ref{qBT2x})
\ba
&& 2 (\ln {\cal V})_x = W_t^- - W_t^+   +{e^{-\Theta}\over 2}(a e^{W^+ + W^- \over 2} - \bar a e^{-W^+ - W^- \over 2})\cr
&& 2 (\ln {\cal V})_x = W_t^+ - W_t^- +{e^{\Theta}\over 2}(-a e^{-W^+ - W^- \over 2} + \bar a e^{W^+ + W^- \over 2}).
\ea
Compatibility conditions of the latter equations lead to the following equation
\be
W_t^+ - W_t^- = {a\over 2} \cosh({W^+ + W^- \over 2} -\Theta)- {\bar a\over 2} \cosh({W^+ + W^- \over 2} +\Theta).
\ee

It is worth emphasizing once more that the equations derived above as $t$ and $x$ evolutions of the defect degrees of freedom are not satisfied simultaneously as it happens in the usual B\"{a}cklund transformation frame. To conclude in the continuum case the $x$ and $t$ relations separately are precisely the same with $x$ and $t$ part of the B\"{a}cklund transformation relations, but still they are not simultaneously satisfied, hence the appellation {\it quasi B\"{a}cklund} transformations still holds. Nevertheless one can suitably define the derivatives at $x \to x_0^{\pm}$ for the left and right theories, but still the defect is frozen at $x=x_0$, this is a subtle process which will be discussed in more detail elsewhere, together with the solution of the derived quasi B\"{a}cklund transformations and the corresponding behavior of the defect degrees of freedom.

\section{Comments}

Several general comments are in order here. As realized the integrable discontinuity may be seen as a B\"{a}cklund transformation fixed at $x_0$ (space-like) or at $t_0$ (time-like). On the other hand B\"{a}cklund transformations as is well known produce (multi)solitonic solutions in non-linear integrable PDEs and also relate different solutions of the integrable PDEs \cite{wahlq, matveev}. Here the picture is somehow modified given that the bulk left and right theories posses solitonic solutions and the defect acts as a transmitting ``wall'' allowing it to interact with  left/right solitonic solutions. Basically in the picture emerging here we are mostly interested in the time (or space) evolution of the degrees of freedom given certain left and right solutions. This is turn can provide information on how the defect interacts with the solitons of the bulk theories. This depiction is arguably more in tune with the quantum description, where the soliton type quantum excitations are transmitted through the defect. The transmission amplitudes may be then explicitly derived either via the Faddeev-Zamolodchikov algebra \cite{delmusi, konle, corrigan-qu1, corrigan-quantum} especially in the frame of integrable quantum field theories or the Bethe ansatz formulation \cite{doikou-quantum} in discrete integrable systems.

It is worth noting however that the complete physical and mathematical picture is not yet fully understood. Significant insight has been gained from the study of discrete integrable models in the presence of local defects, albeit in some sense they are more complicated than the continuous models given that analyticity conditions are missing, thus as has been clear in the text the time evolution equations become non-local and more complicated to deal with. An important issue to consider is the composition of several single soliton Darboux matrices to provide a kind of an extended defect. In fact, the composition of a large number (thermodynamic limit $N \to \infty$) of such fundamental Darboux matrices (see e.g. \cite{matveev}) would be especially interesting to study given that these can be now expressed as integral representations that satisfy a certain version of the Gelfand-Levitan-Marchenko equation. An example of type II defects as a product of type I defects in the particular context of the sine-Gordon here provides such a case, however this needs to be further explored at the level of solutions of the associated differential equation and their asymptotic behavior. At the quantum level composition of multiple Darboux type matrices provide the non-local charges of the underlying quantum algebra. The connection of these objects with the classical analogues is a really interesting question.

Related to the aforementioned questions the key issue one may address is the classical scattering of solitonic excitations. This can be studied by acting on the auxiliary function with various B\"{a}cklund transformations, and then consider the asymptotic behavior of the auxiliary function. Effectively this can be better interpreted by the derivation of the related Gelfand-Levitan-Marchenko equation (see e.g. \cite{FT}) in the presence of local discontinuities via the Zakharov-Shabat dressing (see e.g. \cite{matveev}). Again let us stress that particular emphasis should be given when considering the infinite product of fundamental Darboux matrices acting on the defect point. As already mentioned in the text one could also understand how local Darboux matrices may lead to the quantum analogue of the generating function of the B\"{a}cklund transformation i.e. the $Q$ operator along the lines described in \cite{sklyanin-back}. These are all very interesting as well as intricate issues, which will be addressed in forthcoming investigations.
\\
\\
{\bf Acknowledgements}\\
I am indebted to P. Adamopoulou and G. Papamikos for illuminating discussions and useful comments.


\begin{thebibliography}{1}

\bibitem{delmusi}
G. Delfino, G. Mussardo and P. Simonetti, Phys. Lett. {\bf B328} (1994) 123, {\tt hep-th/9403049};\\
G. Delfino, G. Mussardo and P. Simonetti, Nucl. Phys. {\bf B432} (1994) 518, {\tt hep-th/9409076}.

\bibitem{konle}
R. Konik and A. LeClair, Nucl. Phys {\bf B538} (1999) 587; {\tt hep-th/9793985}.

\bibitem{corrigan-atft}
P. Bowcock, E. Corrigan and C. Zambon, JHEP {\bf 01} (2004) 056, {\tt hep-th/0401020};\\
E. Corrigan and C. Zambon, J. Phys. {\bf A 42} (2009) 304008, {\tt arXiv:0902.1307 [hep-th]}.

\bibitem{corrigan-qu1}
P. Bowcock, E. Corrigan and C. Zambon, JHEP {\bf 08} (2005) 023, {\tt hep-th/0506169}.

\bibitem{corrigan-NLS}
E. Corrigan and C. Zambon, Nonlinearity {\bf 19} (2006) 1447, {\tt nlin/0512038}

\bibitem{corrigan-quantum}
E. Corrigan and C. Zambon, JHEP {\bf 07} (2007) 001, {\tt arXiv:0705.1066 [hep-th]}.

\bibitem{corrigan-fused}
E. Corrigan and C. Zambon, J. Phys. {\bf A 42} (2009) 475203;
{\tt arXiv:0908.3126 [hep-th]};\\
E. Corrigan and C. Zambon, J. Phys. A {\bf 43} (2010) 345201, {\tt arXiv:1006.0939 [hep-th]}.

\bibitem{haku}
I. Habibullin and A. Kundu, Nucl. Phys. {\bf B795} (2008) 549, {\tt arXiv:0709.4611 [hep-th]}.

\bibitem{nemes}
F. Nemes, Int. J. Mod. Phys. {\bf A 25} (2010) 4493; {\tt arXiv:0909.3268 [hep-th]}.

\bibitem{caudrelier}
V. Caudrelier, IJGMMP {\bf 5} No. 7 (2008) 1085, {\tt arXiv:0704.2326 [math-ph]};\\
V. Caudrelier, J. Phys. {\bf A48} (2015) 195203, {\tt arXiv:1411.5171 [math-ph]}.

\bibitem{caudrelier-kundu}
V. Caudrelier and A. Kundu, JHEP 02 (2015), 088, {\tt arXiv:1411.0418 [math-ph]}.

\bibitem{aguire}
A.R. Aguirre, T.R. Araujo, J.F. Gomes, A.H. Zimerman, JHEP, (2011), 12, 56, {\tt arXiv:1110.1589 [nlin.SI]};\\
A.R Aguire,  J. Phys. {\bf A45} (2012) 205205, {\tt arXiv:1111.5249 [math-ph]}.

\bibitem{doikou0}
A. Doikou, Nucl. Phys. {\bf B854} (2012) 153, {\tt arXiv:1106.1602, [hep-th]}.

\bibitem{avan-doikou}
J. Avan and A. Doikou, JHEP 01 (2012) 040, {\tt arXiv:1110.4728 [hep-th]};\\
J. Avan and A. Doikou, JHEP 11 (2012) 008,  {\tt arXiv:1205.1661 [hep-th]}.

\bibitem{doikou-quantum}
A. Doikou and N. Karaiskos, JHEP 02 (2013) 142, {\tt arXiv:1212.0195 [math-ph]};\\
A. Doikou, Nucl. Phys. B877 (2013) 885, {\tt arXiv:1307.2752 [math-ph]}.

\bibitem{robertson}
C. Robertson, J. Phys. {\bf A} 47(2014) 185201, {\tt arXiv:1304.3129, [hep-th]}

\bibitem{ACDK}
J. Avan, V. Caudrelier, A. Doikou, A. Kundu, Nucl. Phys. {\bf B902} (2016) 415,  {\tt arXiv:1510.01173 [hep-th]}.

\bibitem{doikou15}
A. Doikou, {\it A classical variant of the vertex algebra $\&$ the auxiliary linear problem}, {\tt arXiv:1506.08282, [hep-th]}.

\bibitem{wahlq}
H. Wahlquist, in {\it B\"{a}cklund Transformations}, Lect. Notes Math. Vol. 515, pp 162-175.

\bibitem{matveev}
V.B. Matveev and M.A. Salle, {\it Darboux Transformations and Solitons}, Springer Series in non-linear Dynamics, (1991)

\bibitem{Backlund-book}
C. Gu, H. Hu and Z. Zhou, {\it Darboux Transformations in Integrable Systems}, Mathematical Physics Studies, Springer, (2005).

\bibitem{sklyanin-back}
E.K. Sklyanin, CRM Proc. Lecture Notes, 26, Amer. Math. Soc., Providence, RI, (2000).

\bibitem{FT}
L.D. Faddeev and L.A. Takhtakajan, {\it Hamiltonian Methods in the Theory of Solitons},
(1987) Springer-Verlag.

\bibitem{sts}
M.A. Semenov-Tian-Shansky, Funct. Anal. Appl. {\bf 17} (1983), 259.

\bibitem{yang}
C.N. Yang, Phys. Rev. Lett. {\bf 19} (1967) 1312.


\end{thebibliography}
\end{document}